\documentstyle[12pt,aasms4]{article}

\slugcomment{   }

\lefthead{Wiedemann et al.}
\righthead{Search for Methane in the $\tau$ Boo spectrum}

\newcommand{\micronp}{${\mu}m$}
\begin{document}

\title{A SENSITIVE SEARCH FOR METHANE IN THE INFRARED SPECTRUM OF $\tau$ BOOTIS}

\author{G\"{u}enter Wiedemann\altaffilmark{1,3}, Drake Deming\altaffilmark{2}, 
and Gordon Bjoraker\altaffilmark{2,3}} 

\altaffiltext{1}{European Southern Observatory, Karl-Schwarzschild-Str
 2, Garching b. Muenchen, Germany} 
\altaffiltext{2}{Planetary Systems
 Branch, Code 693, Goddard Space Flight Center, Greenbelt MD 20771}
\altaffiltext{3}{Visiting Astronomer at the Infrared Telescope
 Facility (IRTF), which is operated by the University of Hawaii, under
 contract with the National Aeronautics and Space Administration.}
\begin{abstract}

We have searched for a methane signature in the infrared spectrum of
$\tau$ Bootis, produced by the planetary companion.  The observations
comprise 598 low-noise (S/N $\sim$ 100), high resolution ($\lambda /
\delta\lambda = 4 \times 10^4$) spectra near 3044 cm$^{-1}$, which we
analyze by cross-correlating with a modeled planetary spectrum based
on the work of Burrows and Sharp (1999), and Sudarsky et
al. (2000). The $3\sigma$ random noise level of our analysis is $\sim
6 \times 10^{-5}$ stellar continuum flux units, and the confusion
noise limit - measuring the resemblence of a cross-correlation feature
to the spectrum of methane - is $\sim 2.5 \times 10^{-4}$.

We find a significant cross-correlation amplitude of $\sim 3.3 \times
10^{-4}$ continuum units at a velocity near that of the star.  This is
likely due to methane from a low-mass companion in a long-period
orbit.  Fischer, Butler and Marcy (2000) report a long-term velocity
drift indicative of such a companion.  But the system is known to be a
visual binary with an eccentric orbit, and is rapidly approaching
periastron.  Whether the visual companion can account for our
observations and the Fischer et al. velocity drift depends on knowing
the orbit more precisely.  The stability of planetary orbits in this
system also depends crucially on the properties of the binary orbit.

A second cross-correlation feature, weaker and much more diffuse, has
intensity amplitude $\sim 2 \times 10^{-4}$ continuum units and occurs
at a velocity amplitude of $71\ (\pm10)$ km sec$^{-1}$, in agreement
with the orbit claimed for the planet by Cameron et al. (1999).  Like
the first feature, it has passed several tests designed to reject
systematic errors.  We discuss the possibility that this second feature
is due to the planet.

\end{abstract}

\keywords{planetary systems - stars:\ individual (HR 5185) - stars:\
binary (ADS 9025)\ - infrared:\ spectra}

\section{Introduction}

The discovery of a planetary companion to 51 Pegasus (Mayor and Queloz
1995) has been followed by a flood of planet detections using the
Doppler reflex technique (Marcy and Butler 1998, 2000).  The so-called `hot
Jupiters' or `roasters' (Sudarsky et al. 2000), are gas-giants
(Charbonneau et al. 2000) in very short period orbits, where they
both reflect and absorb significant amounts of stellar radiation.
This opens the possibility of detecting their reflection of
the visible stellar spectrum (Charbonneau et al. 1999, Cameron et
al. 1999), or the infrared (IR) spectrum produced by the planets themselves.

The $\tau$ Boo planet is believed to be the hottest of the `roaster'
class, with $T \sim 1600$ Kelvins (Seager and Sasselov 1999, Sudarsky
et al. 2000).  Charbonneau et al. (1999) placed an upper limit on the
reflected flux of $5 \times 10^{-5}$ relative to the star.  However,
Cameron et al. (1999) claim a detection of reflected flux at the $2
\times 10^{-4}$ level, at a planetary radial velocity amplitude of 74
km sec$^{-1}$.  They mention the possibility that planetary molecular
signatures might be detectable.  This paper reports evidence for
methane features in the $\tau$ Boo system, obtained using an IR
spectral deconvolution technique.

\section{Observations}

Observations of $\tau$ Boo were made on UT 29 March - 3 April 1998
using the CSHELL spectrometer (Greene et al. 1993) on the 3-meter NASA
IRTF on Mauna Kea.  More than 1200 spectral frames were acquired under
uniformly excellent observing conditions, and the actual integration
time on $\tau$ Boo totaled to over 31 hours.  Figure 1 shows the
distribution of observations relative to the expected value of the
planetary radial velocity during the run.  The spectral region chosen
was near 3044 cm$^{-1}$, where methane lines occur whose lower state
energy will be significantly populated at the temperature of the
$\tau$ Boo planet.  Numerous weak terrestrial absorptions due to
ozone, methane, and water vapor also occur in this window.  Figure 2
shows a high-resolution solar spectrum bracketing this region from the
McMath-Pierce FTS (Delbouille et al. 1981); it reveals primarily
terrestrial absorption.  A weak solar iron line occurs near 3044.53
cm$^{-1}$ (Livingston and Wallace 1991), but this line should be even 
weaker in the $\tau$ Boo spectrum.  We convolved the FTS spectrum with
a Gaussian whose FWHM was matched to the CSHELL resolution (0.076
cm$^{-1}$), and the resultant simulated CSHELL spectrum is also shown
on Figure 2.

Observations of $\tau$ Boo were typically made as 90 second
integrations at two positions on the slit, by nodding the telescope
between `a' and `b' positions.  The slit was oriented East-West, and
the width was 0.5 arc-sec. Multiple reads of the 256x256 InSb array
were made to reduce read noise to negligible levels. We typically sum
four spectral frames in the order `abba' with the sign of the `b'
frames made negative to subtract the background.  Observations of a
continuum lamp were made before, during, and after each night, and
used to flat-field the spectra.  A Krypton emission lamp was observed
in various grating orders, and used to establish the wavelength
calibration.  Each of the 303 sums shows the star at both `a' and `b'
slit positions.  Extracting both spectra from each sum yields a total
of 598 spectra, after a few low-quality spectra are rejected.  A
sample spectrum, extracted by the processing described in Sec. 4, is
shown on Figure 2.  All of the features visible in these spectra are
due to terrestrial absorption.  The observed $\tau$ Boo spectra are
similar to the convolved FTS spectrum, the differences being
attributable to humidity and altitude. The water line at 3042.39
cm$^{-1}$, for example, is much weaker in the $\tau$ Boo spectrum from
Mauna Kea.  Lunar spectra were also acquired with CSHELL to further
characterize the terrestrial absorption spectrum.

\section{Model Template Spectra}

One form of spectral deconvolution is to cross-correlate (CC) the
observed spectra with a modeled template spectrum (Deming et
al. 2000).  We computed a methane template from a model atmosphere, and
also used it to simulate the planet in a set of `synthetic planet'
spectra (see Sec 4). The template spectrum was assumed to be
independent of orbital phase. We used the temperature versus pressure
profile for the `class IV' atmosphere shown by Sudarsky et al. (2000,
their Figure 1).  Number fractions for methane at each pressure were
calculated using the formulae given by Burrows and Sharp (1999), and
solar abundances.  A constant value was used for the continuous
opacity, chosen to produce unit optical depth where the temperature
equaled the effective temperature (1620K).  The emergent flux spectrum
was calculated, at sub-Doppler resolution, by a quadrature integration
over emergent angle of intensity spectra.  The line opacity profiles
were taken as Voigt profiles, with depth-dependent Doppler broadening
and damping.  The damping was adopted to be 0.1 cm$^{-1}$ per bar.
The most current line data for methane were used (Brown 1999).  A
$T^{3/2}$ dependence was adopted for the rotational partition function
(Robiette and Dang-Nhu 1979), and the vibrational partition function
was calculated from the band centers listed by Brown et al. (1989).
The emergent flux spectrum was convolved to the CSHELL spectral
resolution.

Figure 3 shows the resultant methane spectrum; it is similar to the 51
Peg planet spectrum as modeled by Goukenleuque et al. (2000). Although
most carbon will be in the form of CO at this temperature (Fegley and
Lodders 1996, Burrows and Sharp 1999), the methane absorption in this
spectral region remains sufficiently strong that the thermal
modulation is limited by the overlapping profiles of saturated lines.
In our template model, optical depth unity in the methane lines occurs
in a pressure range of 0.01 to 0.1 bars, versus 10 bars for the
continuum.  A model with methane fractions reduced by an
order-of-magnitude was also computed (Figure 3), and it shows
substantially more modulation than does the nominal model.

\section{Data Processing}

All of our data processing and analysis used procedures specifically
written for this problem. The first stage in the data processing was to
flat-field each frame, remove cosmic ray defects, and `hot pixels'.
Flat-fielding consisted of dividing each frame by a continuum lamp
frame.  Cosmic rays produce very large deviations, and stellar spectra
which were directly affected by them were discarded.  Cosmic ray
defects in the background were repaired in a manner similar to the hot
pixels, which were replaced by an average of surrounding pixels.
Following these corrections, the frames are summed in groups of four
(`abba').

At the second stage of the data processing, we created a set of
synthetic planetary spectra, by adding a Doppler-shifted and
intensity-scaled version of the methane template spectrum to the
flat-fielded data sums. The Doppler shifts were based on the Cameron
et al. (1999) velocity amplitude, with the orbital ephemeris from
Butler et al. (1997) and Marcy (1998).  The intensity scale factor
(ratio of planetary to stellar continuum fluxes) was $6.7 \times
10^{-4}$.  This follows from a planetary radius of $1.8$ Jupiter radii
(Cameron et al. 1999) and 1620 Kelvin effective temperature (Sudarsky
et al. 2000).  The stellar temperature is near 6450 Kelvins (Perrin et
al. 1977, Ford et al. 1999), with rotational $v\sin i = 15$ km
sec$^{-1}$ (Baliunas et al. 1997).  We adopted a stellar radius of
$2.0R_{\bigodot}$, which results from the Cameron et al. (1999)
inclination ($29^{\circ}$), combined with the assumption that the
planetary orbit is locked to the stellar rotation.  The synthetic and
real spectra were processed and analyzed in parallel, using identical
procedures.

Some spectra showed an intermittent fringing effect, not removed by
the normal flat-fielding procedure.  Fortunately, we were able to
verify (by comparing continuum lamp scans made at different times)
that the spacing of the fringes was constant.  We therefore removed
them using a Fourier notch filter; we applied exactly the same filter
to the synthetic planet data.  The slight tilt of the slit was
corrected by spline-interpolating intensities in each row onto a
standard wavenumber scale.  We computed a wavenumber calibration for
each row using six lines from Krypton lamp spectra taken on the first
night, and applied zero-point corrections for subsequent nights using
atmospheric lines.  The wavenumber versus pixel relation used both a
linear and quadratic term, as needed for optimal approximation of the
grating equation.  For each spectrum, we calculated the average
spatial profile along the slit at both the `a' and `b' positions, and
these profiles were fit to the intensities at each wavelength using
linear least-squares.  This is equivalent to the optimal extraction
algorithm discussed by Horne (1986).

Terrestrial atmospheric lines were removed by scaling an atmospheric
template spectrum in intensity, using linear least-squares.  Our
atmospheric template was constructed by averaging all of the
continuum-removed `a' and `b' spectra, producing a template for each
set.  We also explored using lunar spectra to construct atmospheric
templates, but the number of lunar spectra is far fewer, and we
obtained the lowest CC noise levels by averaging the $\tau$ Boo
spectra themselves.  Because our observations cover a range of phase
over nearly two orbital periods, planetary lines will not reinforce in this
averaging, so the loss of planetary signal by fitting and subtracting
this template will be minimal.  Moreover, this and other processing
effects are evaluated by the effect on the synthetic spectra, which are
treated in an identical fashion.

After all processing, the resultant signal-to-noise (S/N) ratios
(typically $\sim 100$) are close to the photon statistical limit, as
shown for the collection of `b' spectra on Figure 4 (the `a' spectra
are of equal quality).  The remaining differences between the measured
noise and the theoretical limit are due to variability of a few
atmospheric lines relative to the template, as well as small
intermittent imperfections in the array detector which are not
completely removed by flat-fielding.

\section{Cross-Correlation Analysis}

Following Deming et al. (2000), we look for the presence of the planet
from significant peaks in a CC function, $C(v_{a},v_{g})$, where
$v_{a}$ is the amplitude of the planetary orbital radial velocity, and
$v_{g}$ is the geocentric radial velocity of the center of mass (for
our purpose, the stellar radial velocity).  We derive CC error
estimates using a Monte-Carlo technique (see Sec. 6). Although $v_{g}$
is known (-19 km sec$^{-1}$ for our observations), we treat it as a
free parameter, to help distinguish between real and false
signals.  Similarly, we vary $v_{a}$ over a range which includes
negative amplitudes, out of phase with plausible planetary signals. We
compute $C(v_{a},v_{g})$ as:
\begin{equation}
C(v_{a},v_{g})=N^{-1}\sum_{i=1}^N [\frac{\sum_{j=1}^M r_{ij} \tau_{ij} w_{ij} }
{(\sum_{j=1}^M \mid \tau_{ij} \mid)(M^{-1}\sum_{j=1}^M w_{ij})}]
\end{equation}

where $r_{ij}$ is the residual intensity in $\tau$ Boo spectrum $i$ at
wavenumber $\nu_{j}$, in units of the stellar continuum, after the
processing described in Sec. 4.  $N$ is the number of spectra (598),
and $M$ is the number of wavelengths (256). The template values
$\tau_{ij}$ are computed as:
\begin{equation}
\tau_{ij}=F_{\nu}^{\prime}(\nu_{j}(1-\frac{v_{g}+v_{p}^{i}}{c}))
\end{equation}

where $F_{\nu}(\nu_{j})$ is the planetary flux, in units of the stellar
continuum flux, at wavenumber $\nu_{j}$ in the frame of the planet, and
the prime indicates that the average value over the observed spectral
interval has been subtracted.  The $\frac{v_{g}+v_{p}}{c}$ factor
shifts the wavenumber scale to the observers frame. The planetary
radial velocity $v_{p}^{i}$ is computed as:
\begin{equation}
v_{p}^{i}=v_{a} sin(2\pi (t_{i}-t_{0})/P)
\end{equation}

where $t_{i}-t_{0}$ is the time since 1998, March 29.58179 UT, and
$P=3.31267$ days (Butler et al., 1997, Marcy, 1998).  The denominator
in Eq. (1) provides a normalization for $C(v_{a},v_{g})$ such that its
peak values are equal to the average amplitude of the planetary lines,
in units of the stellar continuum flux.  Positive CC values correspond
to methane absorption spectra, and negative to emission.  The $w_{ij}$
are weights which allow for the variable quality of the data.  We
currently use:
\begin{equation}
w_{ij}=\sigma_{i}^{-2} \sigma_{j}^{-2}
\end{equation}

where $\sigma_{i}$ is the noise level of spectrum $i$ and
$\sigma_{j}$ is the standard deviation of $r_{ij}$ values at a given
$j$.

\section{Results}

The number of photons detected over all wavelengths and summed over
all 598 spectra is $\sim 2.2 \times 10^{9}$.  Our best-case detection
limit will approach the inverse square-root of this number, $2.1 \times
10^{-5}$.  We replaced the real $r_{ij}$ values with gaussian random
noise whose amplitudes were equal to the photon noise level in each
spectrum, and verified that the noise in $C(v_{a},v_{g})$ agrees
with this limit.  The noise levels achieved in the weighted real data
produce $\sigma = 2.4 \times 10^{-5}$ in $C(v_{a},v_{g})$.  The
computed $C(v_{a},v_{g})$ values are shown as false-color images over
a range of $v_{a}$ and $v_{g}$ in Figure 5.  Two special locations marked on
these images are the position of the center of mass, at $v_{g}=-19$ km
sec$^{-1}$, and the Cameron et al. (1999) claimed amplitude for the
planet at $v_{a}=+74$ km sec$^{-1}$.  Figure 5a shows the results for
the real data, 5b the synthetic data (which equals the real data plus a
synthetic signal), 5c is the result when the data are replaced by
gaussian random noise whose amplitude equals the noise level in each
spectrum, and 5d is the real data cross-correlated with a perturbed
template (described below).  To enhance the visibility of the
planetary signature in Figure 5b, the synthetic signal is multiplied by 10.

The strongest peak on Figure 5a occurs at $v_{g}=-10$ km sec$^{-1}$ and
$v_{a}=-8$ km sec$^{-1}$.  Given that our spectral resolution is $7.5$
km sec$^{-1}$, and template errors are also likely, this peak
coincides with the stellar position to within the errors, and we refer
to it as the `stellar peak' (SP).  Its amplitude is $3.3 \times
10^{-4}$, which is well above the random noise level.  It is present
when both the `a' and `b' spectra are analyzed separately. Since our
template for terrestrial atmospheric removal uses an average of all
the $\tau$ Boo data, this will tend to remove features near
$v_{a}=0$.  However, the atmospheric template is used with a variable
multiplier (depending on air mass), which will allow some fraction
of features fixed in the stellar frame to survive the atmospheric
corrections.

A second region of enhanced CC amplitude runs from the middle left to
upper right, and passes close to the signal from the synthetic planet (5b)
at the Cameron et al. (1999) amplitude of $74$ km sec$^{-1}$.  We
refer to this region as the `planetary streak' (PS).  Its amplitude is
$\sim 2 \times 10^{-4}$ in stellar continuum units, also well above
the random noise.  In our analysis, a feature at constant wavelength
in the data will produce such slanted regions of enhanced CC
amplitude. This occurs because a given wavelength is overlapped by a
given template point over many pairs of values in $(v_{a}, v_{g})$.
Noise features can produce slanted patterns in either of two
orthogonal directions.  Planetary signals should be more concentrated
at single points in the $(v_{a}, v_{g})$ plane, depending on the
distribution of the observations in phase, as well as potential
phase-dependent spectral variations.  Since the majority of our
observations ($63\%$) happen to be made at negative planetary radial
velocities (see Figure 1), using too negative a value for $v_{g}$ is partially
compensated by using a smaller $v_{a}$, and the observed CC amplitudes
should slant from upper right to lower left, qualitatively consistent
with the observed slanting.  Overexposed images of the synthetic planet CC
amplitudes show this effect, but to a much smaller degree than the
observed PS.  Like the SP, the PS is present in both the `a' and `b'
data.

We have subjected our results to numerous tests designed to identify
false signals. Cool stellar spectra show vibrational OH lines in this
region (Hinkle et al. 1995), but nothing resembling the SP/PS
structure is produced by repeating our analysis using an OH template.
A hot, optically thin planetary template, made by reducing the column
density in the model atmosphere by several orders of magnitude, causes
both the SP and PS to weaken greatly.  We repeated the CC analysis
using templates wherein the methane line wavelengths in the planetary
model atmosphere calculation are given gaussian random perturbations
with a standard deviation of $1.5$ cm$^{-1}$.  Computing 50 of these
CC images, each representing $\sim 500$ CC values, we find that the
standard deviation of the resultant `confusion noise' is $\sigma = 8.2
\times 10^{-5}$.  A randomly-selected example of such a perturbed CC
analysis is shown on Figure 5d.  Based on this test, the SP is a
$4\sigma$ feature, and the PS is $2.4\sigma$.  We cannot envision how
any error in our analysis could `know' the real wavelengths of the
methane lines in our template (Figure 3), so the appearance of the SP
only with the correct template is credible evidence of a real methane
signature.

Although the perturbed templates test indicates that the SP is a real
methane signature, we must further establish that it is not produced
by terrestrial methane, via an incomplete removal of the atmospheric
spectrum.  We computed several CC images where we deliberately
under-corrected for terrestrial lines.  These images show an increased
noise level, but they do not show any SP/PS enhancement.  We tried
using the terrestrial spectrum (Figure 2) as an analysis template.
This produces a feature similar to the SP but weaker (0.00026), and
slightly closer to the terrestrial rest frame ($v_{a}=-5\ $km
sec$^{-1}$).  So at least a small portion of the terrestrial spectrum
has survived our data processing.  However, the fact that the SP
strengthens and shifts toward the stellar velocity when using the
Figure 3 spectrum as a template indicates a significant contribution
from methane in the $\tau$ Boo system.

\section{Discussion}

Based on the tests described above, we regard the SP as evidence for
methane in $\tau$ Boo. The SP cannot be attributed to the planet in a
3.3-day orbit, because the SP velocity amplitude is near zero (within
the errors).  It cannot be due to the F7 star, which is much too hot
to contain significant quantities of methane. Detailed interpretation
of such a weak feature is necessarily problematic.  Nevertheless, we
tried templates characteristic of different temperatures, but found
insufficient sensitivity to meaningfully constrain the temperature of
the absorbing region(s).  

We considered the possibility that this methane absorption might be
circumstellar.  But methane is not commonly seen in interstellar
regions (Knacke et al. 1985), and is easily destroyed by stellar
ultraviolet radiation (Mount and Moos 1978).  Also, the SP is much
weaker for optically thin templates, so it does not seem indicative of
absorption by a thin homogenous medium (e.g., a tenuous circumstellar
cloud).  Instead, it is more consistent with methane occurring in an
optically thick clump of column density $\sim 10^{21}$ cm$^{-2}$, with
small filling factor relative to the star.  This type of distribution
suggests a compact object, such as a methane-dwarf.  Can there be an
undetected sub-stellar object in the $\tau$ Boo system?  $\tau$ Boo
was discovered to be a visual binary by Otto Struve in 1849, and
Charbonneau (2000) has suggested that the M-dwarf companion may be
contributing to our spectra.  We here discuss the possibility that the
M-dwarf, or a much cooler compact object, is producing the methane signal
which we observe.

The M-dwarf companion is about 7 or 8 visual magnitudes fainter than
the primary, and its spectral classification (M2) is at least 44 years
old (Eggen 1956, attributes it to Kuiper).  Two orbits for the system
have been published, by Hale (1994) and Popovic and Pavlovic (1996).
The orbital elements differ substantially.  For example, Hale (1994)
gives the eccentricity as 0.91 and the period as 2000 years, whereas
Popovic and Pavlovic (1996) derive $e=0.42$ and $P=389$ years.
Although the Hale orbit was published first, Hale made and utilized an
observation of the system in 1991, which was apparently not known to
Popovic and Pavlovic, who used observations prior to 1970. The orbits
indicate that the position angle of the visual companion at the time
of our observations was $30^{\circ}$ (Hale orbit) or $37^{\circ}$ (Popovic and
Pavlovic orbit), with separation near 2.6 arc-sec in both cases.
These values place the companion sufficiently off the CSHELL slit that
no contribution to our observations would be possible.

However, there are additional relevant data for this system.
Responding to our inquiry, Fischer, Butler and Marcy (2000) report
that the primary star exhibits a long-term ($\sim 1990$ to current)
velocity drift of -17 meters sec$^{-1}$ year$^{-1}$, requiring a
companion in a long-period orbit.  Can this reflex motion be caused by
the M-dwarf?  The Hale orbit predicts that the radial velocity of the
M-dwarf will be changing by about 0.2 km sec$^{-1}$ year$^{-1}$ near
periastron (in 2017).  Assuming a mass ratio of $\sim 10$, this would
account for the drift reported by Fischer et al., but only if
periastron occurs near 2000, instead of 2017.  Interestingly, an
earlier periastron would place the M-dwarf close to the slit during
our observations.  If the spectral type is significantly in error
(which is plausible), and the companion is much cooler than M2, then
it might show sufficient strength in the strong $3.3\ \micron$ methane
band to account for the SP in our data.  Also, if the Hale orbit is
correct, the periastron distance of $\sim 22$ a.u. seems to preclude
stable orbits for additional companions with periods of tens of years
(Holman and Wiegert 1999).  Alternately, if the Popovic and Pavlovic
(1996) orbit is correct, the M-dwarf did not contribute to our
observations, nor can it account for the velocity drift observed
by Fischer et al. (2000).  So in this case another companion in the
system would be indicated, and the periastron distance of $\sim 66$
a.u. for the M-dwarf might well permit stable orbits having periods of
tens of years.

As for any signal so small ($3$ parts in $10^{4}$), the reality of the
SP should be confirmed by additional observations.  Moreover, as discussed
above, the interpretation of our data and the Fischer et al. (2000)
velocity drift as indicative of additional sub-stellar companions in
this system, depends crucially on precise knowledge of the orbit for
the M-dwarf. Additional observations of the M-dwarf are badly needed,
including separation, position angle, an updated spectral type, and
observations of infrared colors.

The PS is also well above the random noise level of our analysis, but
only marginally above the `confusion noise' ($2.4\sigma$).  Because
the confusion noise is estimated by applying random perturbations to
the wavelengths of the methane template lines (Sec. 6), it measures not
only the reality of a cross-correlation feature, but also the degree to
which a feature can be attributed to methane.  Accepting the $2.4\sigma$
statistical significance for the moment, we consider the possibility
that this methane signal is produced by the planet.  The amplitude of
the CC peak from the synthetic planet is $3 \times 10^{-5}$
(the peak on Figure 5b has been amplified by 10, for clarity).  So the
PS (at $2 \times 10^{-4}$) is more than six times as large as the
synthetic planet, which we have modeled to be consistent with the
Cameron et al. (1999) planetary parameters.  Since the Cameron et
al. radius is already in modest excess of theoretical values (Guillot
et al. 1996), it may seem that the PS amplitude is implausibly large.
But the amplitude of our synthetic planet signal has been strongly
reduced by the overlap of saturated lines (Figure 2).  The chemical
equilibrium abundance of methane (Burrows and Sharp 1999) will be
competitive with CO only in the coolest regions of the planet's
atmosphere.  If methane is convected to greater depths, then
disequilibrium effects could result in a lower methane abundance
(Griffith and Yelle 1999), and our synthetic planet signal would
increase.  Also, we have assumed a stellar radius of $2R_{\bigodot}$,
which follows from the Cameron et al. orbital inclination, the
rotation data (Baliunas et al. 1997), and a synchronous assumption.  A
more conventional radius for the star (e.g., $1.4R_{\bigodot}$, Ford
et al. 1999) will increase the size of the synthetic planet signal,
since it is measured in stellar continuum flux units.  We conclude
that the amplitude of the PS is plausible for the planet.

At $v_{g}=-19$ km sec$^{-1}$, the PS occurs at $v_{a}=+61$ km
sec$^{-1}$, whereas the Cameron et al. value is $v_{a}=+74$ km
sec$^{-1}$. If we entertain the possibility that the SP is displaced
from the stellar position primarily by template error, then it is more
appropriate to measure the difference in the Figure 5a ordinate between
the SP and PS, which is $+81$ km sec$^{-1}$.  A reasonable estimate
and error range for the PS position is to bracket these absolute and
relative determinations, giving $+71\ (\pm10)$ km sec$^{-1}$. While
this is consistent with the Cameron et al. value, the large spread of
the PS into an extended streak seems inconsistent with a planetary
origin.  Nevertheless, it is possible that strong phase effects in the
planetary spectrum could cause such large spreading in an analysis of
this type.  Observations with cryogenic cross-dispersed echelle
spectrometers on $10-$meter class telescopes (McLean et al. 1998)
should be able to clarify the reality of this feature, and the
role of spectral variation with phase, by providing very high
S/N spectra over a much larger wavelength range.

\section{Acknowledgments}

We thank Adam Burrows for sending us methane cross-section data, Geoff
Marcy for ephemeris data, and Linda Brown for providing the latest
methane line parameters.  We also thank Debra Fischer, Paul Butler and
Geoff Marcy for communicating their results in advance of publication,
and allowing us to quote them, and Bill Hartkopf for discussing the
current status of the visual binary observations.  We acknowledge
extensive discussions on the data analysis and interpretation with
Dennis Reuter and Don Jennings, and insightful comments by Dave
Charbonneau, Mike Mumma, Sara Seager, and Roger Yelle. This research
was supported by the NASA Origins of Solar Systems program.

\clearpage

\begin{figure}[p]
\vspace{-1.in}
\plotone{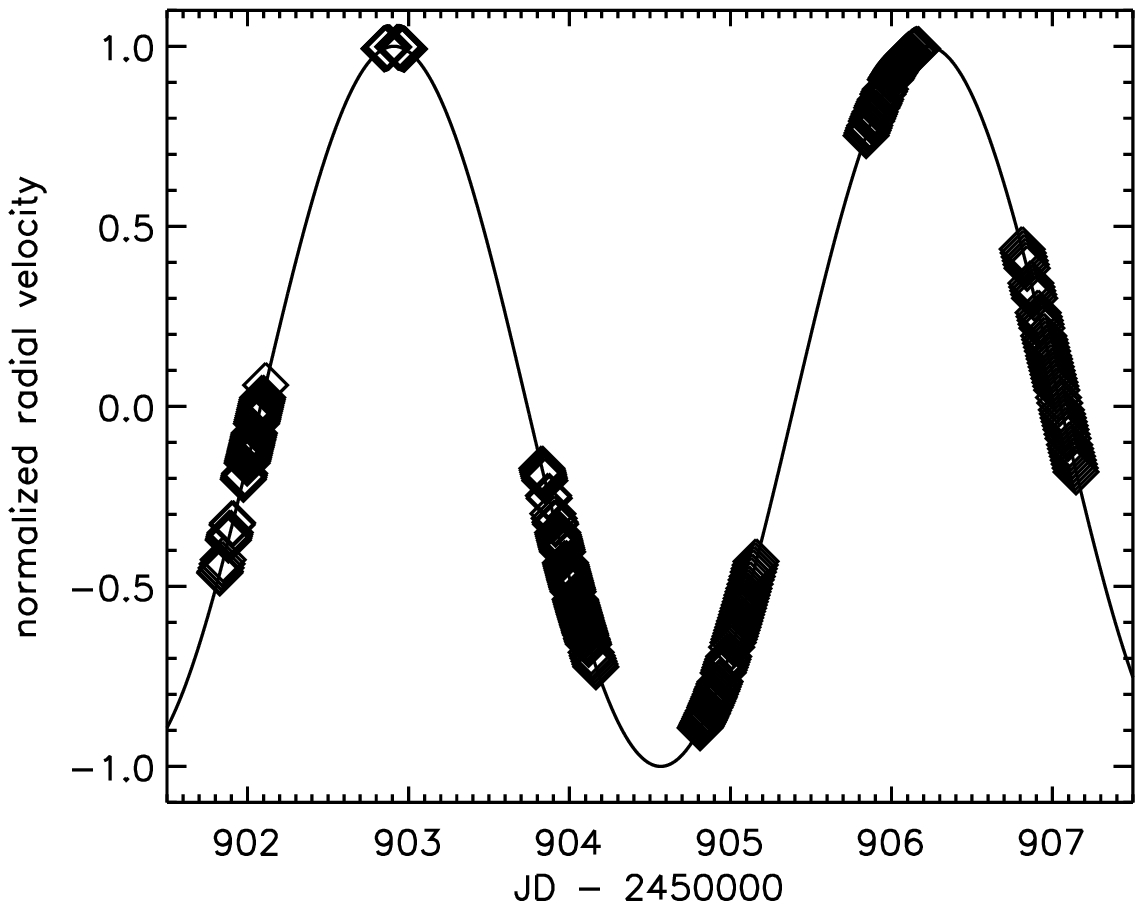}
\caption{Distribution of the observations versus time.  Phase is indicated by
the planetary radial velocity, normalized to amplitude unity.  \label{fig1}}
\end{figure}

\begin{figure}[p]
\vspace{-1.in}
\plotone{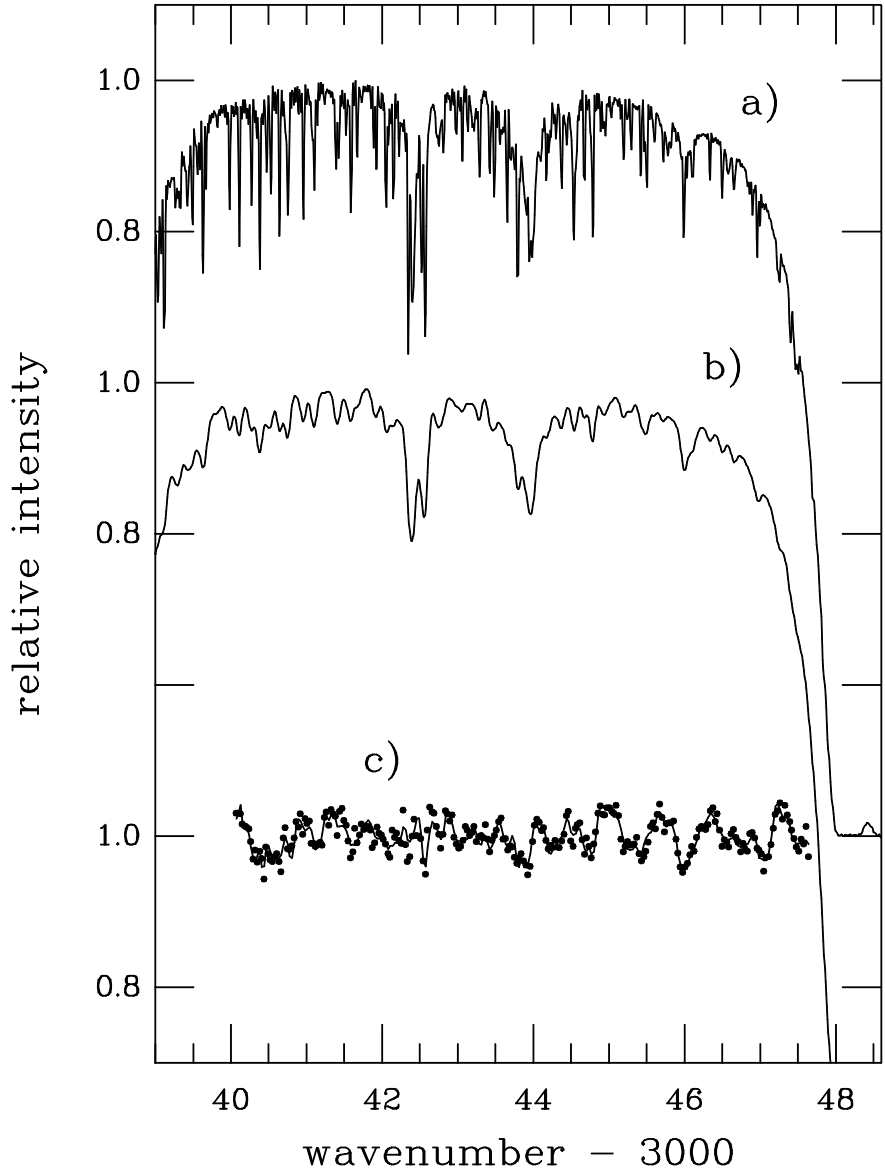}
\vspace{-1.in}
\caption{Sample spectra near 3044 cm$^{-1}$: a) high-resolution FTS
solar spectrum from Kitt Peak, b) FTS solar spectrum convolved to
CSHELL resolution, c) sample CSHELL spectrum (points) with the
template fit (line). \label{fig2}}
\end{figure}

\begin{figure}[p]
\vspace{-1.in}
\plotone{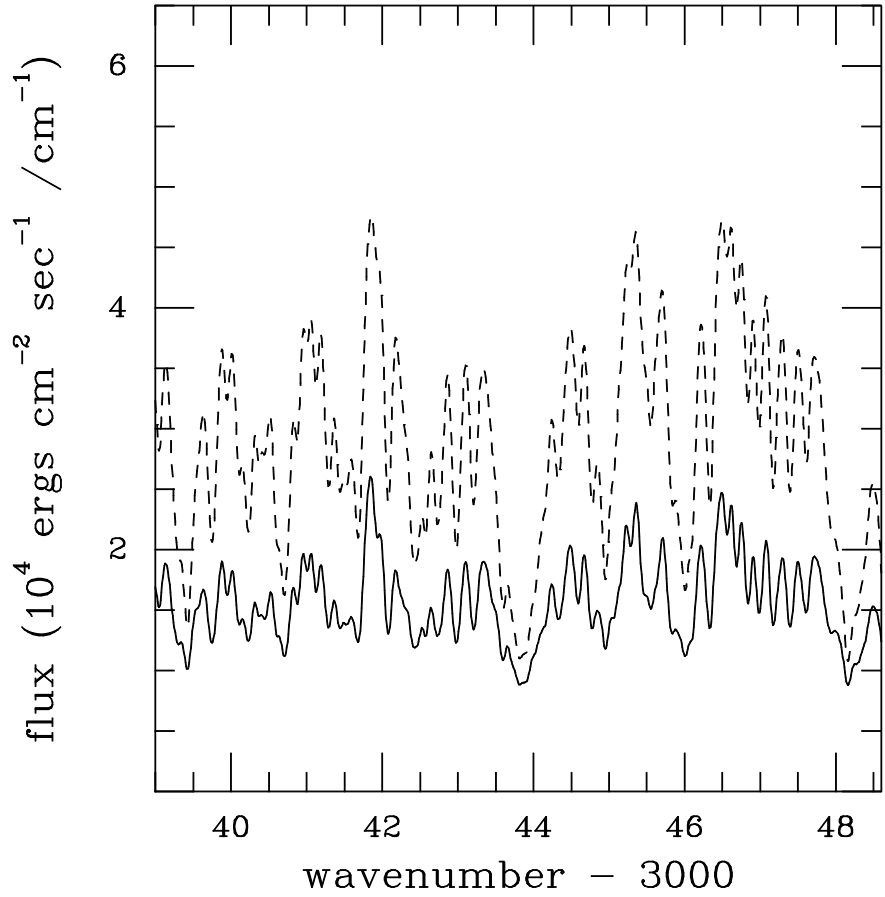}
\vspace{-1.in}
\caption{Modeled methane spectra near 3044 cm$^{-1}$. Solid line is the
nominal $\tau$ Boo planet model, and the dashed line shows the effect
of reducing the methane mixing ratio by a factor of 10. The continuum
level is at $6.5 \times 10^{4}$.  \label{fig3}}
\end{figure}

\begin{figure}[p]
\vspace{-1.in}
\plotone{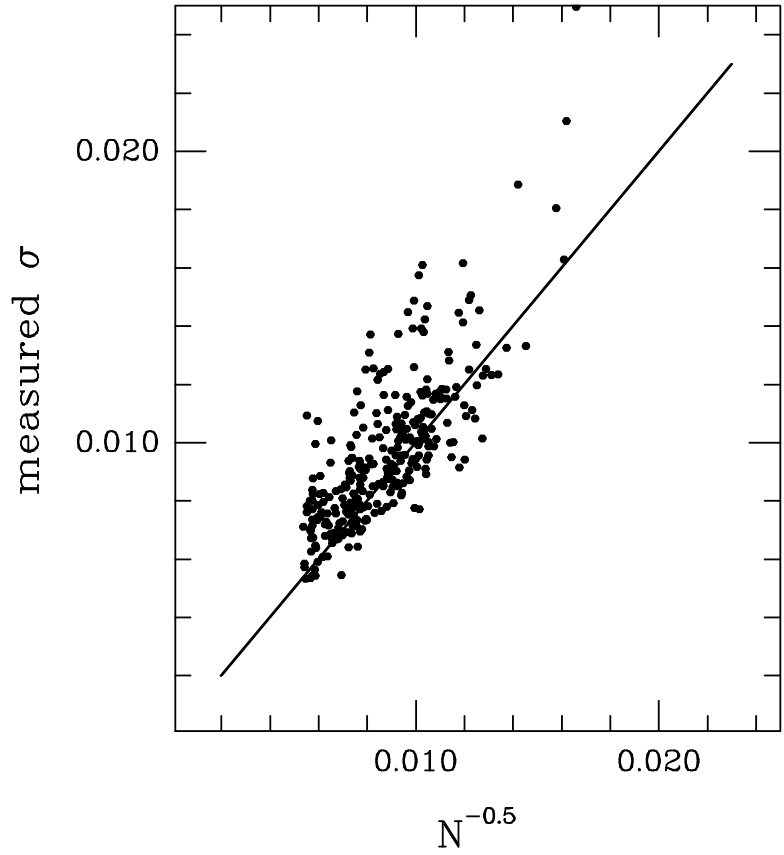}
\vspace{-1.in}
\caption{Measured noise level (standard deviation, in continuum units)
versus the inverse root of the total number of photoelectrons detected
per wavelength point.  The line gives the photon statistical limit. \label{fig4}}
\end{figure}

\end{document}